\documentclass[cernpreprint, atlasdraft=false, texlive=2023, UKenglish, texmf, orcidlogo]{atlasdoc}
\usepackage[subcaption]{atlaspackage}
\usepackage{array}
\usepackage{atlasbiblatex}

\usepackage{atlasphysics}
\graphicspath{{logos/}{figures/}}

\usepackage{ANA-TRIG-2022-03-PAPER-defs}
\usepackage{multirow}
\AtlasTitle{Fast $b$-tagging at the high-level trigger of the ATLAS experiment in LHC Run~3}

\AtlasAbstract{%
The ATLAS experiment relies on real-time hadronic jet reconstruction and $b$-tagging to record fully hadronic events containing $b$-jets.
These algorithms require track reconstruction, which is computationally expensive and could overwhelm the high-level-trigger farm, even at the reduced event rate that passes the ATLAS first stage hardware-based trigger.
In LHC Run~3, ATLAS has mitigated these computational demands by introducing a fast neural-network--based $b$-tagger, which acts as a low-precision filter using input from hadronic jets and tracks. It runs after a hardware trigger and before the remaining high-level-trigger reconstruction.
This design relies on the negligible cost of neural-network inference as compared to track reconstruction, and the cost reduction from limiting tracking to specific regions of the detector.
In the case of Standard Model $HH \rightarrow b\bar{b}b\bar{b}$, a key signature relying on $b$-jet triggers, the filter lowers the input rate to the remaining high-level trigger by a factor of five at the small cost of reducing the overall signal efficiency by roughly 2\%.

}

\PreprintIdNumber{CERN-EP-2023-111}

\AtlasJournal{JINST}

\AtlasJournalRef{JINST 18 (2023) 001 P11006}
\AtlasDOI{10.1088/1748-0221/18/11/P11006}

\hypersetup{pdftitle={ATLAS document},pdfauthor={The ATLAS Collaboration}}

\begin{document}

\maketitle

\tableofcontents

\section{Introduction}
%


%
%
%
%
%
%
%
%
%
%
%
%
%
%
%
%
%
%
%
%
%
%

The ATLAS experiment at the LHC relies on selective triggers to capture events containing \bhadron-initiated jets (\bjets), which are associated with a variety of physics processes. Within the Standard Model these processes range from frequently occurring top quark production (predominantly decaying into a \bquark and a $W$ boson) to rare processes like associated production of a Higgs boson with top quark pairs (where all particles decay hadronically), or Higgs pair production (where at least one of the Higgs bosons decays into a $b\bar{b}$ pair~\cite{HIGG-2019-04,HIGG-2020-14}).
Beyond the Standard Model many theories feature decays of hypothetical new particles into final states containing \bquarks~\cite{HDBS-2018-41, SUSY-2019-12}.

Extensive studies were conducted by the LHC experiments on the properties of the Higgs boson, discovered in 2012~\cite{HIGG-2021-23,CMS-HIG-22-001}, revealing no evidence of physics beyond the Standard Model so far.
Of particular interest is the trilinear Higgs boson self-coupling, $\lambda_{HHH}$, which connects the Higgs boson mass to the vacuum expectation value and plays a role in the production of Higgs boson pairs, offering valuable tests for the electroweak symmetry breaking mechanism.
Deviations from the Standard Model prediction of the Higgs boson self-coupling are pertinent in theories extending the Standard Model and as such measuring this coupling is a key goal of the LHC physics programme.
Nevertheless, detecting Higgs boson pairs at the LHC presents challenges due to their low production cross-section: strong evidence is only expected to emerge during the HL-LHC data taking~\cite{Dainese:2703572, ATL-PHYS-PUB-2022-053}.
To effectively investigate this process it becomes crucial to optimise the trigger efficiency for Higgs boson pair production.
Among the various decay channels of $HH$, the one with the highest branching ratio involves both bosons decaying into two $b$-quarks each.
This specific channel poses significant triggering difficulties as there are few discernible features in the event beyond the presence of the $b$-quarks.

Jets originating from \bquarks produce
\bhadrons, which have a non-negligible lifetime, on the order of
$10^{-12}$~s~\cite{Zyla:2020zbs}.
These \bhadrons can travel a measurable distance, on the order of 2 mm, before decaying, leaving the
striking signature of a \newterm{secondary} decay vertex separate from the \newterm{primary} interaction vertex of the proton--proton ($pp$) collision.
In addition to the decay length, \bjets can be distinguished from jets
originating from gluons or light quarks (i.e.,~jets that do not contain a heavy-flavour
$b$- or $c$-hadron, called \newterm{light-jets}), by the larger charged-particle multiplicity, the high
fraction of jet energy carried by tracks displaced from the primary interaction
vertex, and the invariant mass of these displaced tracks.

Algorithms for identifying, or \newterm{tagging}, \bjets take as input the properties of individual jets,
together with well-reconstructed tracks within these jets.
Optimisation of these algorithms aims at minimising the background rate, or equivalently, maximizing the rejection of light-jets
and $c$-jets (i.e. jets originating from $c$-hadrons) for a fixed target \btagging efficiency.
The algorithms are separately optimised for use at the trigger level and in
offline reconstruction, due to the tight CPU and memory constraints associated with the online
processing and the need for maximum precision offline.
Remaining within these limitations while balancing background
rejection and signal efficiency is the main challenge of the \bjet trigger.

The \bjet trigger selections were among the most CPU-intensive selections at the trigger
level during Run~2 data taking~\cite{ATL-DAQ-PUB-2016-002}
and as such any reduction in the time to process a single \bjet event can free considerable resources.
These free CPU cycles can then be rededicated to improve trigger level reconstruction and expand the physics reach of the experiment.

A novel approach to \btagging at trigger level was introduced for Run~3 data taking to minimise excess computation with minimal cost in signal acceptance.
This was done by implementing a rapid \btagging preselection, which is executed after identifying jets using calorimetry but before the computationally expensive full-event track reconstruction. The preselection reduces tracking to specific detector areas around high-energetic jets, effectively serving as an early background rejection tool. This rejection method decreases the processing load from multi-$b$ signatures, resulting in a reduction of CPU usage and keeping overall resource use within the capacity of the computing farm's budget.

The new approach is presented in this paper, which is structured as follows. A
brief description of the ATLAS detector and trigger system is given in
Section~\ref{sec:detector}. The algorithms used as input to \bjet trigger
selections are described in Section~\ref{sec:triggerAlgos}. The new \newterm{fast
\btagging} is described in Section~\ref{sec:fastBtagging} and its performance in
Section~\ref{sec:perf}. The paper concludes with a summary in Section~\ref{sec:conclusion}.
Details on the ATLAS Run~3 trigger, including tracking CPU consumption and timing measurements, as well as an assessment of the overall performance of the $b$-tagging at the trigger level, are provided in Ref.~\cite{trigger_run_3}.


%

%
%
\section{The ATLAS detector and trigger system in LHC Run~3}
\label{sec:detector}


The ATLAS detector~\cite{ATLAS:2023dns,PERF-2007-01} covers nearly the entire solid angle around the collision point.\footnote{ATLAS uses a right-handed coordinate system with its origin at the nominal interaction point (IP) in the centre of the detector and the \(z\)-axis along the beam pipe. The \(x\)-axis points from the IP to the centre of the LHC ring, and the \(y\)-axis points upwards.
Cylindrical coordinates \((r,\phi)\) are used in the transverse plane, \(\phi\) being the azimuthal angle around the \(z\)-axis.
The pseudorapidity is defined in terms of the polar angle \(\theta\) as \(\eta = -\ln \tan(\theta/2)\). Angular distance is measured in units of \(\Delta R \equiv \sqrt{(\Delta\eta)^{2} + (\Delta\phi)^{2}}\).}
It consists of an inner tracking detector surrounded by a thin superconducting solenoid, electromagnetic and hadronic calorimeters, and a muon spectrometer incorporating three large superconducting toroidal magnets. The tracking detector and the calorimeters are relevant components to this paper and they are summarised below.

The inner detector system is immersed in a \SI{2}{\tesla} axial magnetic field and provides charged-particle tracking in the range \(|\eta| < 2.5\).
The high-granularity silicon pixel detector covers the vertex region and typically provides four measurements per track, the first hit being normally in the insertable B-layer installed before Run~2~\cite{ATLAS-TDR-2010-19,PIX-2018-001}.
It is followed by the semiconductor tracker (SCT), which usually provides eight measurements per track.
These silicon detectors are complemented by the transition radiation tracker (TRT), which enables radially extended track reconstruction up to \(|\eta| = 2.0\).
The TRT also provides electron identification information based on the fraction of hits above a higher energy-deposit threshold corresponding to transition radiation.

The calorimeter system covers the pseudorapidity range \(|\eta| < 4.9\).
Within the region \(|\eta| < 3.2\), electromagnetic calorimetry is provided by barrel and endcap high-granularity lead/liquid-argon (LAr) calorimeters, with an additional thin LAr presampler covering \(|\eta| < 1.8\) to correct for energy loss in material upstream of the calorimeters.
Hadronic calorimetry is provided by the steel/scintillator-tile calorimeter, segmented into three barrel structures within \(|\eta| < 1.7\), and two copper/LAr hadronic endcap calorimeters.
The solid angle coverage is completed with forward copper/LAr and tungsten/LAr calorimeter modules optimised for electromagnetic and hadronic measurements, respectively.

The ATLAS trigger and data acquisition system is responsible for the online processing and event selection for permanent storage and offline analysis. It employs a two-level trigger system~\cite{TRIG-2016-01} to select data at an average rate of about 3~kHz. Its first level (Level-1 or L1) is hardware based. It uses custom-made electronics to select events after processing signals coming from the calorimeters and muon chambers. The Level-1 system runs at a fixed latency of 2.5~$\mu$s and accepts events at a maximum rate of 100~kHz. The second level of the trigger system, the high-level trigger (HLT), uses a dedicated computing farm of approximately 60,000 physical processing cores or 2.0M HS06~\cite{Michelotto:2010zz} to run algorithms similar to those used in the offline reconstruction. The HLT software was redesigned for Run~3 to support multi-threaded execution, allowing for more efficient use of computing resources. %
The algorithms execution order is optimised to run fast algorithms first, providing early background rejection, and more precise and CPU-intensive algorithms later to make the final event selection.
Both the L1 and HLT triggers, in addition to performing selection, can identify Regions-of-Interest (RoIs) in $\eta$, $\phi$, and $z$, which limit the regions of the detector on which subsequent algorithms are executed.
The processing uses data from all detectors either in RoIs only or in more extended $\eta$-$\phi$ and $z$ range (up to the full detector), depending on the algorithms. The HLT makes a decision within a few hundred milliseconds on average. The selected events are sent to permanent storage with a throughput of up to 8~GB/s.

An extensive software suite~\cite{ATL-SOFT-PUB-2021-001} is used in data simulation, in the reconstruction
and analysis of real and simulated data, in detector operations, and in the trigger and data acquisition
system of the experiment.


%

\section{Algorithms for \bjet trigger}
\label{sec:triggerAlgos}


The algorithms running at the HLT are split into two general types: reconstruction algorithms, which create objects such as tracks, clusters and jets; and hypothesis algorithms, which apply selections on the reconstructed objects, or combinations of those.
Three categories of algorithms are relevant to the work presented in this document: inner detector tracking, jet finding and \bjet finding.  The related reconstruction algorithms are described in what follows.

\subsection{Inner detector tracking}

A previous paper describes the design and performance of the HLT inner detector tracking on ATLAS during the LHC Run~2~\cite{TRIG-2019-03}.
The algorithms are improved for Run~3 but the Run~2 description is sufficient for the purposes of this paper.
The tracking algorithms operate in two steps: a \newterm{fast tracking} stage typically followed by a \newterm{precision tracking} stage.

Before any tracking algorithm is run, the data in the silicon modules of the inner detector are reconstructed into clusters. %
This can be run on the full detector or within RoIs; the later approach saves both processing time and data transfer bandwidth. The TRT data within RoIs is only processed for the precision tracking stage. In events characterised by large jet multiplicities or large number of simultaneous $pp$ interactions per bunch crossing (pile-up), multiple individual RoI constituents can be merged into a \newterm{super-RoI}, such that the tracking algorithm does not run multiple times in RoIs with large overlapping regions, which would lead to wasted CPU resources and biases due to track double-counting.

The fast track finder (FTF) provides track candidates to use early in the trigger or when the resource-intensive precision tracking is not affordable.
In the FTF pattern recognition %
initial track candidates are formed using a simple track finding algorithm that extends the track candidates into further layers to find additional hits.
The FTF design focuses on efficiency over purity.
As described further in Section~\ref{sec:trigJetFinding}, full-event fast track finding is a prerequisite for the standard Run~3 trigger jet reconstruction.
In addition these tracks are used to construct a primary vertex, defined as the vertex with the highest scalar sum of the squared transverse momenta ($\pT^2$) of all associated tracks.

In the case of \bjet triggers, FTF tracking runs again, with a lower minimum \pT requirement and looser requirements on track quality, in RoIs in restricted $\eta$, $\phi$ and $z$ regions around \bjet candidates.
These FTF tracks are used to seed the precision tracking, which is used for the final \bjet selection in the trigger.
Precision tracking takes the FTF tracks as input and applies a version of the offline tracking algorithms configured to run fast in the trigger. This includes algorithms that improve the purity of the FTF tracks by removing track duplicates. The track candidates are extended into the TRT to improve the momentum resolution.
Overall, the trigger tracking efficiency is driven by the FTF one, since these tracks are used as seeds to the precision step. The precision tracking performs higher quality fits that improve the purity and the quality, approaching that of the offline reconstructed tracks.

\subsection{Jet finding}
\label{sec:trigJetFinding}
Jet finding at trigger level is described in the ATLAS Run~2 trigger commissioning paper~\cite{TRIG-2016-01}, and is summarised only briefly here.
At Level-1, jets are defined as $4 \times 4$ or $8 \times 8$ trigger tower windows for which the summed electromagnetic and hadronic transverse energy exceeds predefined thresholds and which surround a $2 \times 2$ trigger tower core that is a local maximum. Besides the transverse energy thresholds, Level-1 selections depend on the multiplicity or the topology~\cite{TRIG-2019-02} of jets in the event.
Typically, jet algorithms are executed at the HLT on events accepted by Level-1 jet trigger selections.

A major improvement in the Run~3 jet trigger is the introduction of particle flow, adopting the offline reconstruction algorithms described in Ref.~\cite{PERF-2015-09}.
In this paradigm, jets (referred to as \empflow jets) are clustered from charged and neutral particle flow constituents.
Candidates for \btagging are clustered with the anti-$k_t$ algorithm~\cite{Cacciari:2008gp} with a radius parameter $R=0.4$.
Charged constituents are built from selected tracks and matched to topological clusters~\cite{PERF-2014-07} of calorimeter energy deposits.
To reduce the effects of pile-up, only charged constituents matched to the primary vertex are considered.
Neutral constituents are built from calorimeter energy clusters where the energy associated to the charged constituents is subtracted to avoid double-counting.

While \empflow jets improve energy resolution and reduce the effects of pile-up, the added dependence on tracking adds significant CPU costs.
Jets reconstructed from calorimeter energy clusters alone (referred to as \emtopo jets) are used in the fast \btagging described in Section~\ref{sec:fastBtagging}, to avoid the CPU burden incurred by tracking.
Both \empflow and \emtopo jets are calibrated to improve the fidelity of the momentum of the measured jet with respect to the underlying particle shower, detailed in the ATLAS Run~2 offline jet calibration paper~\cite{JETM-2018-05}.
In the case of \emtopo jets, trigger reconstruction uses a variant of the Run~2 offline calibration that uses calorimeter information only.

\subsection{\bjet finding}

The HLT \bjet finding in ATLAS uses a combination of multivariate algorithms that typically take as input features the reconstructed jets, reconstructed tracks, and the position of the primary vertex.
In Run~3, ATLAS unified HLT algorithms with their offline counterparts.
Previous ATLAS papers describe the Run~2 \bjet trigger~\cite{TRIG-2018-08}, and the updated algorithms that form the basis of Run~3 offline and trigger \btagging~\cite{FTAG-2019-07}.

In Run~2, fast tracks reconstructed with the FTF algorithm within super-RoIs were used for primary vertex finding; precision tracks were reconstructed within \emtopo jets that passed a minimum transverse energy (\ET) threshold.
The final stage of the \bjet trigger assessed the probability that these jets originated from a \bhadron decay.
This probability was evaluated in two steps. The first step used low-level algorithms that matched tracks to jets, reconstructed secondary vertices, and identified tracks with large impact parameters relative to the primary vertex. The second step, which ran on the output of these low-level algorithms, employed machine learning algorithms that provided excellent discrimination between \bjets and light-jets or $c$-jets.

Since in Run~2 the \bjet finding involved running precision tracking in all jet RoIs above a minimum \ET{} threshold, the \bjet trigger selections were the most CPU-intensive selections at the ATLAS HLT\@.
To reduce resource demands, tighter jet \ET{} and $\eta$ selections can be applied before tracking is executed, but this reduces the physics acceptance of the resulting triggers.

\begin{figure}[h]
\centering
\includegraphics[width=0.8\textwidth]{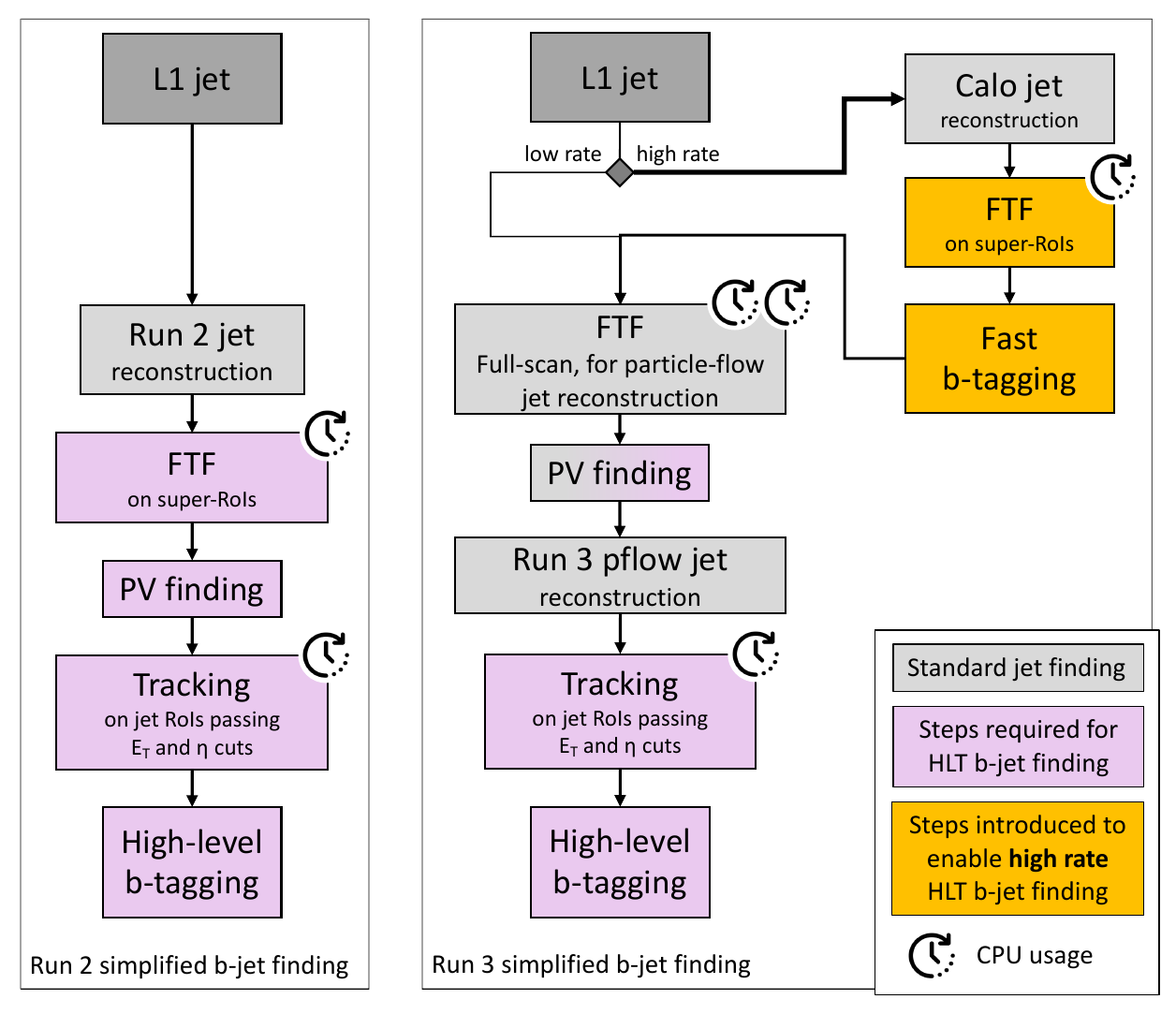}
\caption{Simplified schematic descriptions of the \bjet trigger selections in two different ATLAS trigger implementations: the Run~2 implementation on the left, and the Run~3 implementation on the right.
The Run~3 implementation follows two different execution paths: nearly all high-rate triggers follow the case described here, with preselection for running at high L1 input rate; in cases where the L1 input rate is already acceptably low the preselection is omitted. }
\label{fig:configComparisons}
\end{figure}

A new approach was introduced for Run~3 that significantly reduces the CPU needs for \bjet trigger finding while maintaining excellent efficiency for key physics processes, such as Standard Model \dihfourb{} decays.
This approach exploits the efficacy of machine learning in identifying \bjets of sufficient quality for preselection at the HLT, while using lower quality input objects relative to the state-of-the-art \btagging algorithms developed within ATLAS\@.
The deployment of the \newterm{fast \btagging} is schematically demonstrated in Figure~\ref{fig:configComparisons}, where it is also compared with the Run~2 online \bjet reconstruction.

In this paper, we use the term \newterm{fast $b$-tagging} to describe the newly introduced fast $b$-tagging preselection, which uses the \fastdips{} tagger, presented in the following section.
To simplify matters, we also use the term \fastdips{} to refer to the entire fast $b$-tagging preselection.
Additionally, we refer to \newterm{high-level $b$-tagging} as the final selection used at the HLT, which relies on a tagger in the DL1 series~\cite{FTAG-2019-07}, which is optimised for the trigger.

%


\section{Fast \btagging}
\label{sec:fastBtagging}


The fast \btagging sequence spans three steps: (1) jets are reconstructed from energy deposits in the calorimeter, (2) fast tracking is run within a super-RoI defined by the reconstructed jets, and finally (3) \bjets are identified with \fastdips{}, a deep-sets--based neural network~\cite{NIPS2017_f22e4747}, based on the offline ATLAS DIPS algorithm~\cite{ATL-PHYS-PUB-2020-014}, which ingests tracks and jets as inputs. Events can be rejected following either the jet reconstruction or the \btagging steps. As a low-precision filter, the sequence is optimised to reduce backgrounds without substantially reducing signal efficiency. The three fast \btagging sequence steps are described in the following.

\subsection{Jet reconstruction and tracking}

As introduced in Section~\ref{sec:trigJetFinding}, \emtopo jets are reconstructed and serve as inputs to the first event-level filtering step. A trigger chain can require some number of these jets above a given $\ET$ threshold and within a pseudorapidity range. Events that fail this requirement are rejected.

Each jet that passes the above selection seeds an RoI for track reconstruction. These RoIs cover the entire $z$-range of the interaction region, and project outward in a slice from the beamline, with a $\phi$ and $\eta$ half-width of 0.3. To avoid duplicated track reconstruction from overlapping RoIs, all the RoIs within a single event are merged into one super-RoI, and track finding is run once in the union of the regions.

\subsection{The \btagging algorithm, \fastdips}
\label{sec:fastDips}

Tracks within the super-RoI are associated to the nearest jet if they fall within a cone around the jet\footnote{The standard ATLAS flavour-tagging association cone, defined by $\Delta R < 0.239 + \exp( -1.22 - 0.0164 / \gev \cdot \pt )$, is used here.}, which shrinks with growing jet \pt. The jets are then classified as either $b$- or $c$-hadron initiated, or as light, by a retrained DIPS--based neural network.
In simulated events, ATLAS uses a standard jet flavour definition to match generator-level hadrons with reconstructed jets.
A $b$-hadron within $\Delta R = 0.3$ of the centre of the nearest jet labels the jet as a $b$-jet. Otherwise the same procedure is repeated for $c$-hadrons, then $\tau$-leptons.
Any remaining jets are labelled as light.
Both $b$- and $c$-hadrons must have $\pt > 5\,\GeV$ to be considered for labelling.

Tracks are represented by the variables summarised in Table~\ref{tab:variables}.
Each associated track is fed through a feed-forward neural network, which maps the initial track representation to a 128-dimensional latent space. The full jet representation in this space is computed by summing the latent vectors representing tracks over all tracks in the jet.
The sum of these latent vectors is then mapped to a three-element jet flavour posterior $(p_b, p_c, p_u)$ via another feed-forward network, where the three elements correspond to jets that are \bhadron initiated, $c$-hadron initiated, or any other flavour.
These posteriors are collapsed to a single discriminating variable $D_b$ with the formula:
\begin{equation}
D_b \equiv \log \frac{p_b}{p_c f_c + (1 - f_c) p_u}, %
\end{equation}
where $f_c$ is an adjustable constant used to tune the charm rejection for the tagger. In the results that follow $f_c = 0.018$, the value used for the offline DIPS implementation.
A jet is considered \btagged when the $D_b$ discriminant is above a specific threshold.
A trigger chain can require any number of jets to pass a \btagging requirement: events that fail this requirement are rejected.
\begin{table}[h]
\centering
\begin{tabular}{c | m{0.70\linewidth}}
Variable & Description \\
\hline \hline
\multicolumn{2}{c}{Track Kinematics} \\
\hline
$\dzero$ & Distance of closest approach to the beamline \\
$\dzerolife$ & Lifetime signed $\dzero$ \\
$\zzerobs$ & Displacement between beamspot centre and closest approach to beamline, projected along beamline \\
$\log \zzerobsuncert$ & Log of uncertainty in $\zzerobs$ \\
$\log \left(\ptfrac\right)$ & Log of fraction of jet \pt carried in track \\
$q/p$ & Particle charge divided by momentum \\
\hline
$\deta$ & \multirow{3}{*}{Angular separation between track and jet} \\
$\dphi$ & \\
$\dr$ & \\
\hline \hline
\multicolumn{2}{c}{Track Quality} \\
\hline
$\npix$ & Number of pixel hits \\
$\nsct$ & Number of SCT hits \\
$\ninner$ & Number of innermost pixel layer hits \\
$\nninner$ & Number of next-to-innermost pixel layer hits \\
$\ndof$ & Number of degrees of freedom in track fit \\
$\chisq$ & $\sum_{\text{hits on track}} (r / \sigma_r)^2 \quad[r \equiv \text{hit residual}, \quad \sigma_r \equiv \text{residual uncertainty}]$ \\
\end{tabular}
\caption[Table of DIPS Variables.]{Inputs for
fast \btagging neural network. The conventional \dzero takes the sign of $(\vec{p}_0 \times \dzerovec) \cdot \hat{z}$, where \dzerovec is the track displacement at the closest approach to the beamline and $\vec{p}_0$ is the momentum at that point. A lifetime-signed variant, \dzerolife, is given a positive sign if $|\phi_0 - \phi_{\text{jet}}| < \pi / 2$ and a negative sign otherwise, where $\phi_0$ and $\phi_{\text{jet}}$ are the $\phi$ components of the track displacement and the jet momentum, respectively.
}
\label{tab:variables}
\end{table}

The network was trained on ten million jets from simulated $t\bar{t}$ interactions that include all jet flavours, with hyperparameters similar to those selected for the original \textsc{DIPS} tagger~\cite{ATL-PHYS-PUB-2020-014}.
To prevent the discriminant from preferentially selecting jets based on kinematics, the jets were resampled to have the same two-dimensional $\pt$ and $\eta$ distributions for each flavour.
The discriminant was trained in the \textsc{Keras} interface to \textsc{Tensorflow}~\cite{tensorflow} and ported to the ATLAS trigger software with \textsc{lwtnn}~\cite{lwtnn,network}.


\section{Performance assessment}
\label{sec:perf}


The fast \btagging approach was evaluated in Monte Carlo (MC) simulations before being deployed to the trigger menu for online use. The performance was also evaluated using the first Run 3 collision data. This section provides a summary of the performance assessment for both simulations and data, including the impact on the CPU consumption of the HLT farm. Additionally, this section discusses the effect of this new algorithm on the acceptance of the \dihfourb{} signal.

\paragraph{Metrics used for performance assessment.} Receiver Operating Characteristic (ROC) curves are used to show the dependency of the $b$-jet efficiency to the light-jet rejection, which is the inverse of the light-jet efficiency.  Both quantities are evaluated with simulated \ttbar{} events. For the ROC curves and elsewhere, the efficiency is defined using \emph{true} $b$-jets, defined by the labelling described in Section~\ref{sec:fastDips}.
The $b$-jet efficiency is the ratio of the number of true $b$-jets \btagged with algorithm $X$, over the number of true $b$-jets with no \btagging applied:
\begin{equation} %
b\textrm{-jet efficiency of algorithm $X$}= \frac{\#~\textrm{true $b$-jets tagged with algorithm $X$}}{\# ~\textrm{true $b$-jets}}.
\end{equation}
This also allows comparisons between different preselection methods.
We calculate the light-jet efficiency by applying the same formula to light-jets.
The inverse of this efficiency is defined as the light-jet \newterm{rejection}, and is plotted on the $y$-axis in the ROC curves presented in this paper.

To evaluate the correlations between \fastdips{} and the high-level $b$-tagging algorithm used at the HLT, we define conditional efficiencies. The conditional $b$-jet efficiency of the fast $b$-tagging algorithm \fastdips{} compared with the high-level $b$-tagging is defined as follows:
\begin{equation} %
\textrm{Conditional $b$-jet efficiency} = \frac{\#~\textrm{true $b$-jets tagged with both fast and high-level $b$-tagging}}{\#~\textrm{true $b$-jets tagged with high-level $b$-tagging}}.
\end{equation}

Finally, we evaluate the performance of the fast $b$-jet preselection in both data and MC simulations as a function of the leading trigger-level reconstructed \empflow jet \pt, where the high-level $b$-tagging is applied, to estimate the impact of this additional preselection to the final object of interest.
Two efficiencies are defined, as follows:

\begin{equation}\label{eq:jetEff}
\textrm{Jet-level $b$-tagging efficiency} = \frac{\textrm{\# jets with both a fast and a high-level $b$-tag}}{\textrm{\# jets with a high-level $b$-tag}},
\end{equation}

\begin{equation}\label{eq:evtEff}
\textrm{Event-level $b$-tagging efficiency} = \frac{\textrm{\# events with both $\ge1$ fast and $\ge1$ high-level $b$-tag}}{\textrm{\# events with $\ge1$ high-level $b$-tag}}.
\end{equation}

For the jet-level efficiency, we apply a simple geometric matching between \empflow jet and the \emtopo jet, requiring $\Delta R < 0.3$ and a relative \pT{} difference less than 10\%.
This matching is required to compare the high-level $b$-tagging, which is applied on \empflow jets, to \fastdips{}, which uses \emtopo jets.
The event-level efficiency is determined by counting the events that contain at least one high-level $b$-tag (denominator), and then counting the subset of events with at least one fast $b$-tag (numerator). In the event-level efficiency no geometric matching applied between the two tagged jets.

The studies presented in this paper use trigger-level reconstructed \empflow jets and high-level $b$-tagging as reference quantities instead of offline quantities. This choice was made to evaluate the impact of fast $b$-tagging with respect to an online high-level $b$-tagging algorithm, which is similar to the offline \btagging.

\subsection{Algorithm optimization and performance assessment in MC simulations}

MC simulations were used to train the fast $b$-tagging algorithm and assess its performance before data taking. Simulated $t\bar{t}$ events produced in $pp$ collisions were used to provide a sample of $b$-, $c$- and light-flavour jets. The production of
\ttbar\ events was modelled using the
\POWHEGBOX~v2~\cite{Frixione:2007nw,Nason:2004rx,Frixione:2007vw,Alioli:2010xd}
generator. %
The events were
interfaced to \PYTHIA~8.230~\cite{Sjostrand:2014zea} to model the
parton shower, hadronisation, and underlying event, with parameter
values set according to the A14 tune~\cite{ATL-PHYS-PUB-2014-021} and
using the \nnpdftwo set of PDFs~\cite{Ball:2012cx}. The decays of
bottom and charm hadrons were performed by
\EVTGEN~1.6.0~\cite{Lange:2001uf}.
Similar \btagging algorithms have shown limited simulation dependence, on the order of 10\% for efficiency and background rejection, for common ATLAS event simulation chains~\cite{ATL-PHYS-PUB-2020-009}.

The effect of multiple $pp$ interactions in the same event was
modelled by overlaying the hard-scatter interactions with events from
the \PYTHIA~8.160 generator, using the \nnpdftwo PDF set
and the A3 parameter tune~\cite{ATL-PHYS-PUB-2016-017}.
Particle interactions with the detector
are simulated with ATLAS software~\cite{SOFT-2010-01} based on $\GEANT$~\cite{Agostinelli:2002hh}. Events generated with $\sqrt{s}=13$~\TeV{} were used for training the fast $b$-tagging algorithm, while events generated with $\sqrt{s}=13.6$~\TeV{} were used for performance evaluation.

Since the background trigger rate is driven by light flavour jets, the results presented in this paper focus on light jet rejection and omit $c$-jet rejection, which is lower but otherwise similar as documented in other ATLAS results described above.
Besides ROC curves and efficiency calculations, the CPU consumption was also evaluated and taken into account for deciding the optimal operational points of the algorithm.
These three aspects are discussed in this section.

\begin{figure}[h]
\centering
\subcaptionbox{\label{fig:rocoptpt}}[0.45\textwidth]{
\includegraphics[width=\hsize]{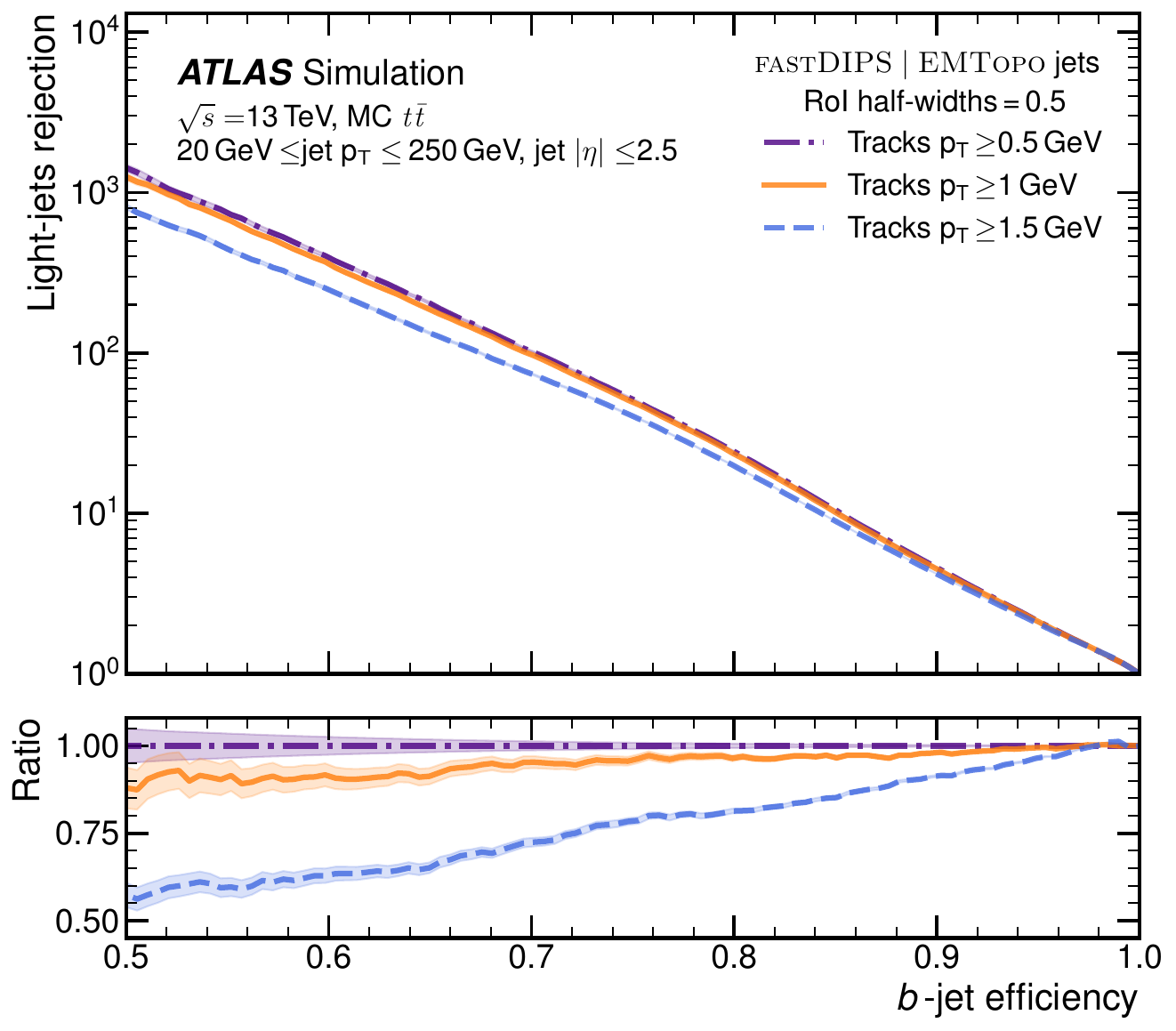}
}
\subcaptionbox{\label{fig:rocoptdr}}[0.45\textwidth]{
\includegraphics[width=\hsize]{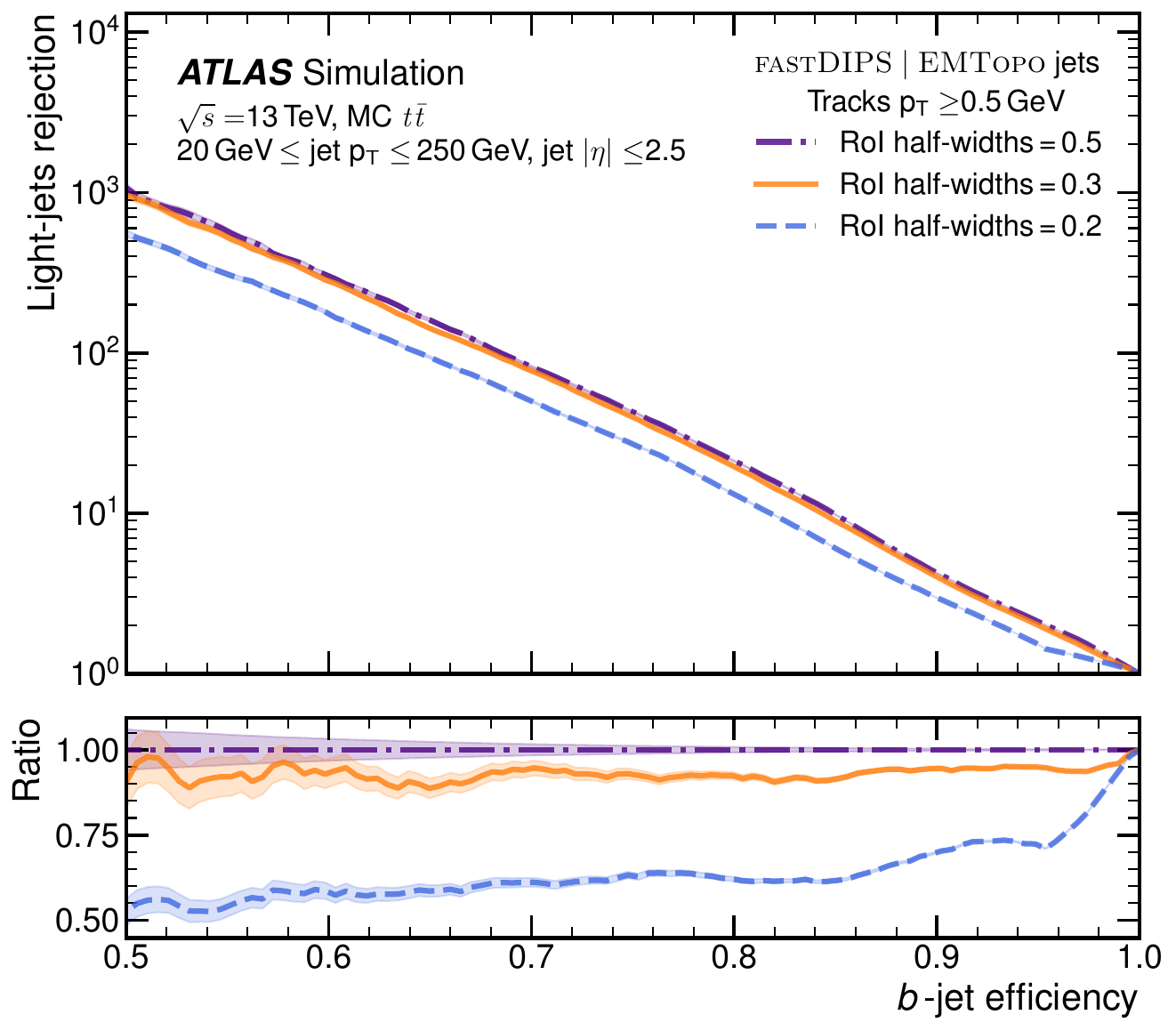}
}
\subcaptionbox{\label{fig:CPUTrackpT}}[0.5\textwidth]{
\includegraphics[width=\hsize]{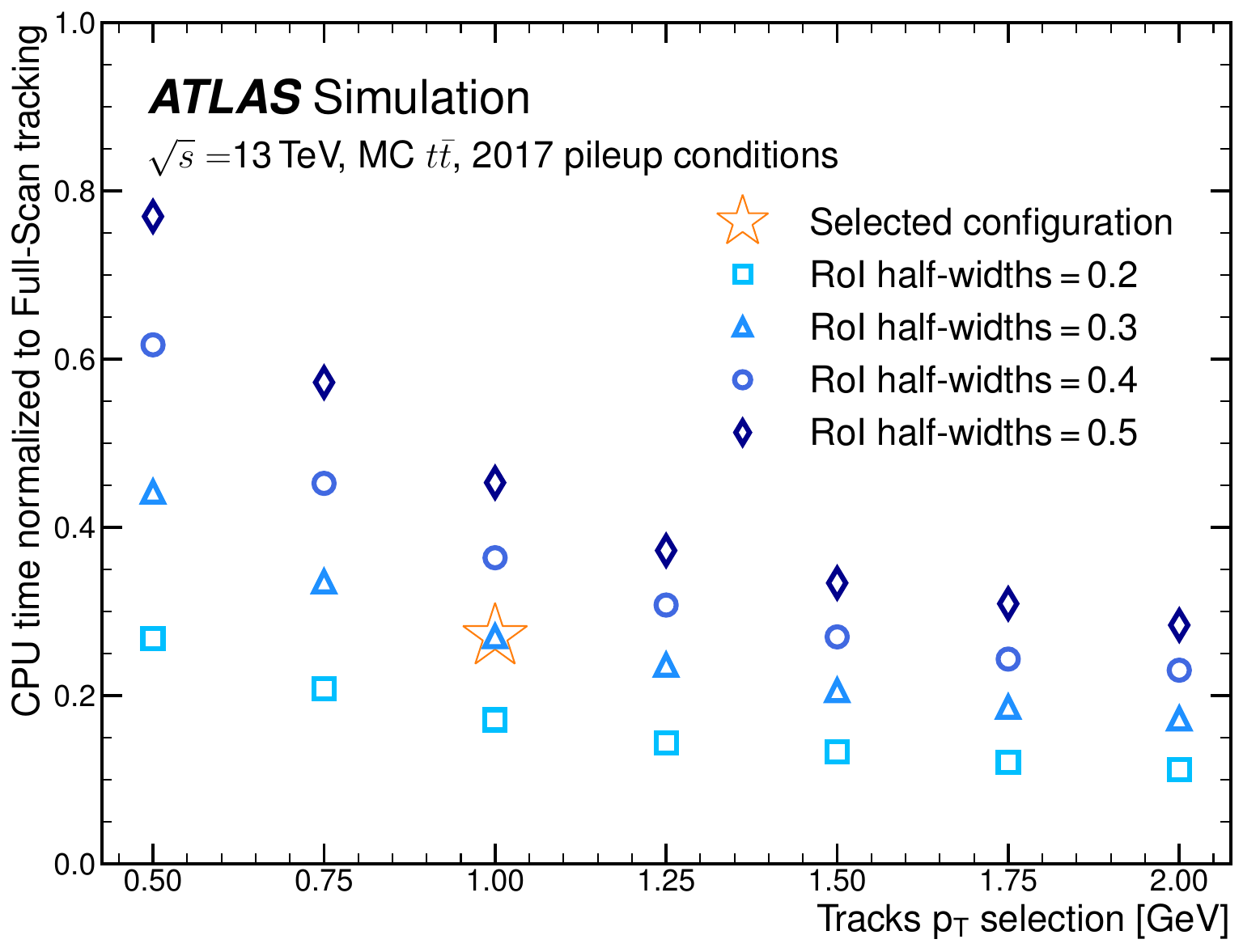}
}
\caption[Optimization of the RoIs used as input for fast track finder execution]{%
Optimization of the RoIs used as input for fast track finder execution.
Figures \subref{fig:rocoptpt} and \subref{fig:rocoptdr} show the light jet rejection ($1/\text{light jet efficiency}$) as a function of the \bjet{} efficiency, for several choices of minimum track \pT{} and RoI size.
Statistical uncertainties for each ROC curve, represented with shaded regions, are computed assuming binomial efficiency errors.
Figure \subref{fig:CPUTrackpT} shows the CPU time required to reconstruct tracks in RoIs of different sizes and centred on the jet axis, normalised to the time employed for running the same tracking algorithms over the full inner detector acceptance.
It demonstrates the relative impact of the algorithm configurations on the CPU requirements.
The full detector tracking requires track $\pt > 1\,\GeV$.
All points, and the full detector baseline CPU estimate, use the same simulated 13~\TeV{} $t\bar{t}$ events, and assume 2017 pile-up condition.
The final configuration selected for Run 3 is indicated in \subref{fig:CPUTrackpT} by an open star.
}
\label{fig:ROC_optimiz}
\end{figure}

The fast $b$-tagging algorithm performance is evaluated for various requirements on the minimum track \pT{}, and for several RoI sizes around the jet axis, given in $\eta$ and $\phi$ half-widths (half of the full width, also discussed in Ref.~\cite{TRIG-2018-08}).
The results are shown in Figure~\ref{fig:ROC_optimiz}.
A minimum \pT{} requirement of 1.0~\GeV{} performs slightly worse than 0.5~\GeV, while a minimum requirement of 1.5~\GeV{} results in a large degradation of \btagging performance.
Only a small loss in performance is observed when decreasing the RoI size from 0.5 to 0.3, but a substantial loss is seen for smaller RoIs.
The CPU time required to reconstruct tracks in a RoI centred on the jet axis is also evaluated and presented in Figure~\ref{fig:CPUTrackpT}.
The final working point for the 2022 data taking is defined at RoI $\eta$ and $\phi$ half-widths of 0.3 and track $\pT \ge 1$~\GeV.
This working point ensures that the CPU cost of fast tracking is less than 30\% of the cost of the same algorithm with full inner detector acceptance; the same fast tracking algorithm runs at full inner detector acceptance in a next step in the trigger processing (as shown in Figure~\ref{fig:configComparisons}), but at lower rate, after the \fastdips{} preselection.

\begin{figure}[h]
\centering
\subcaptionbox{\label{fig:ROCcurves}}[0.49\textwidth]{
\includegraphics[width=\hsize]{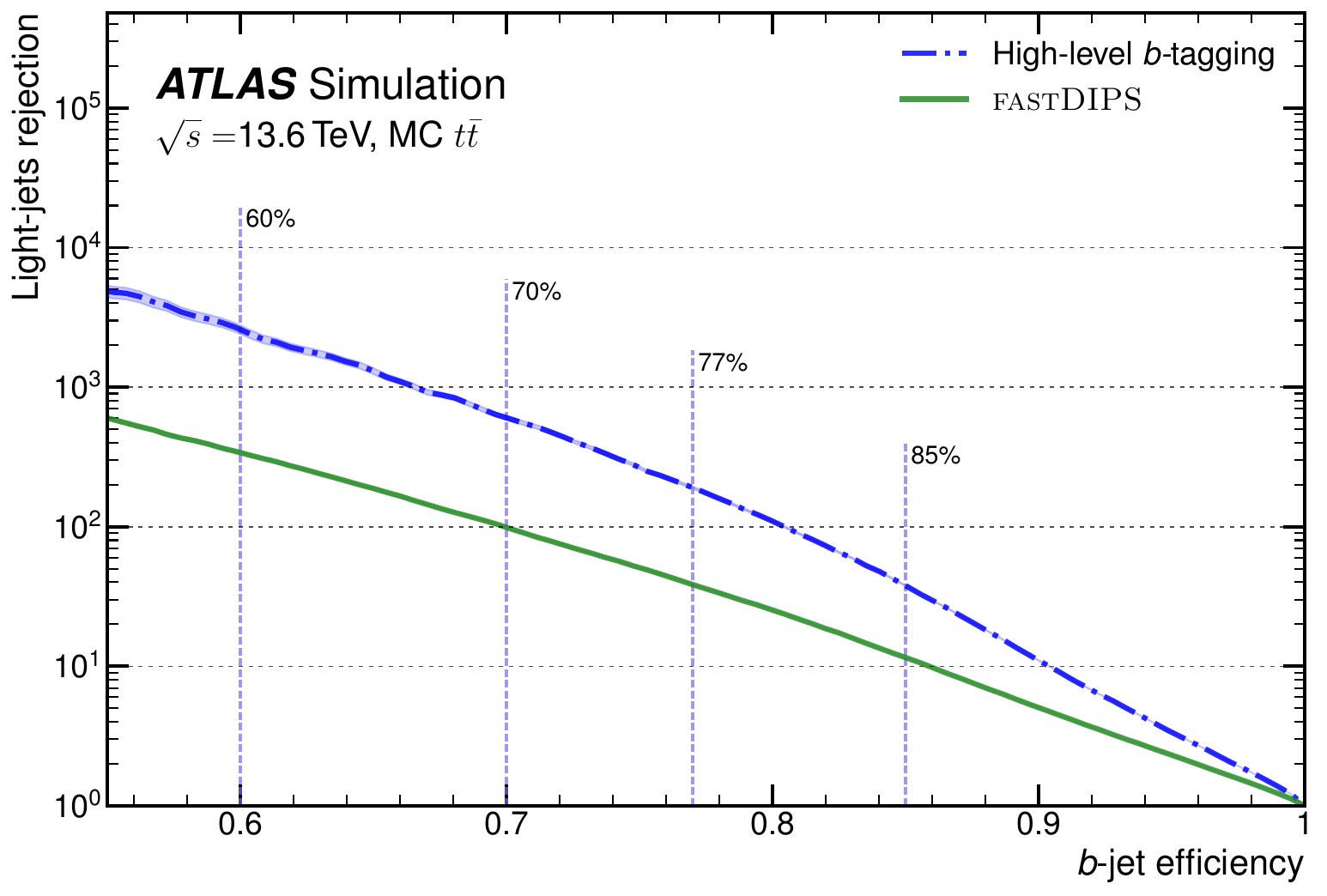}
}
\subcaptionbox{\label{fig:conditionalEfficiencies}}[0.49\textwidth]{
\includegraphics[width=\hsize]{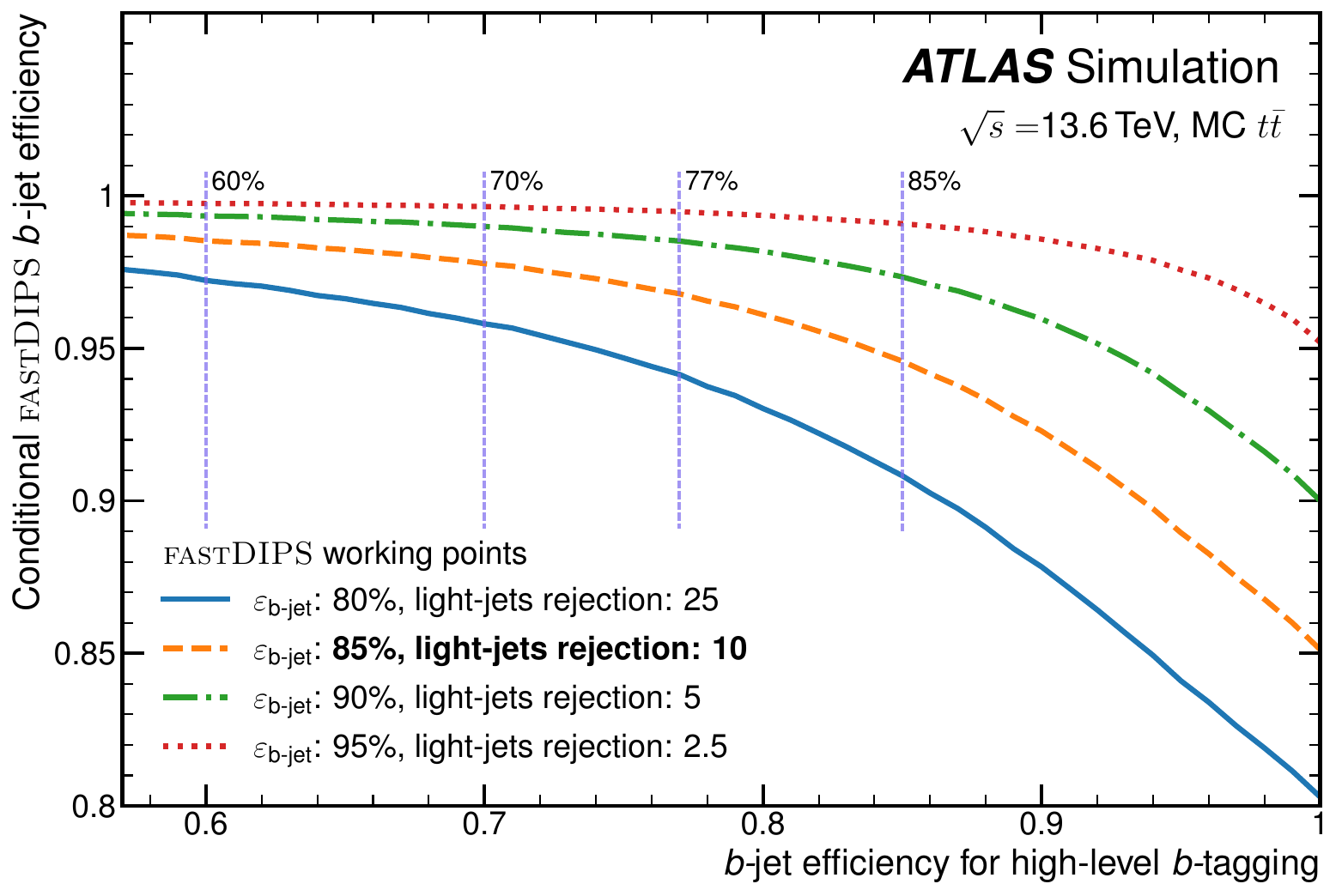}
}
\caption[Comparisons of fast and high-level \btagging]{\label{fig:conditional}%
Comparisons of fast and high-level \btagging.
Fast \btagging is applied to  \emtopo jets, while high-level \btagging is applied to \empflow jets.
Figure \subref{fig:ROCcurves} gives the light-jet rejection as a function of \btagging efficiency for both approaches, fast and high-level $b$-tagging.
Statistical uncertainties for each ROC curve, represented with shaded regions, are computed assuming binomial efficiency errors.
Figure \subref{fig:conditionalEfficiencies} gives the conditional efficiency for fast \btagging to identify a true  \emtopo \bjet,
as a function of the efficiency of the high-level \btagging selection on \empflow jets.
The latter are matched with \emtopo{} jets.
Jets are required to have $\pt > 20\,\GeV$.
The different curves show the conditional efficiencies for multiple \fastdips{} discriminant selections.
The working point with 85\% efficiency is used in the trigger.
The purple vertical dashed lines represent the most common working points used for \btagging.
}
\end{figure}

The fast \btagging performance is compared to the high-level \btagging performance in Figure~\ref{fig:conditional}. A comparison of the light-jet rejection, which drives the rate reduction, is shown in Figure~\ref{fig:ROCcurves}, while Figure~\ref{fig:conditionalEfficiencies} shows the conditional efficiency of \fastdips{} with respect to the high-level $b$-tagging algorithm. For a \fastdips{} preselection with 85\% efficiency, the light jets rejection, and thus rate reduction achieved, is a factor of ten. The conditional efficiency of this fast $b$-tagging preselection with respect to the high-level $b$-tagging algorithm without any preselection is at least 90\% for any chosen working point.

\subsection{Performance assessment in Run~3 collision data}%

To verify that the simulation-based results above accurately describe data, we relied on an event selection that is enriched in $t\bar{t}$ events.
Unless otherwise specified, all objects referenced below are reconstructed in the HLT.
The $t\bar{t}$ events are obtained by requiring the presence of an energetic electron, a muon without any Level-1 requirement, and at least one \empflow jet \btagged at the 80\% working point. %
To ensure a high efficiency for the lepton triggers, further requirements are applied on the offline electrons and offline muons. We require a single \textit{tight} identified electron~\cite{EGAM-2018-01} with $\pt{} > 28$~\GeV{} and  $| \eta |< 2$ and a single muon with $\pt{} > 25$~\GeV{} and  $| \eta | < 2$.
Also, for these studies both \emtopo and \empflow jets are required to be reconstructed with a \pT{} greater than 25~\GeV{} and $| \eta | < 2.5$.
The latter have an additional requirement\footnote{$\text{JVT} \geq 0.5$ for jets with $\pt < 60\,\GeV$, otherwise no requirement.} on the jet vertex tagger score~\cite{PERF-2014-03} to suppress pile-up jets.
The same treatment is performed on the $t\bar{t}$ MC sample that is used for the comparison.

\begin{figure}[h]
\centering
\subcaptionbox{\label{fig:Elvl_ptEff}}[0.49\textwidth]{
\includegraphics[width=\hsize]{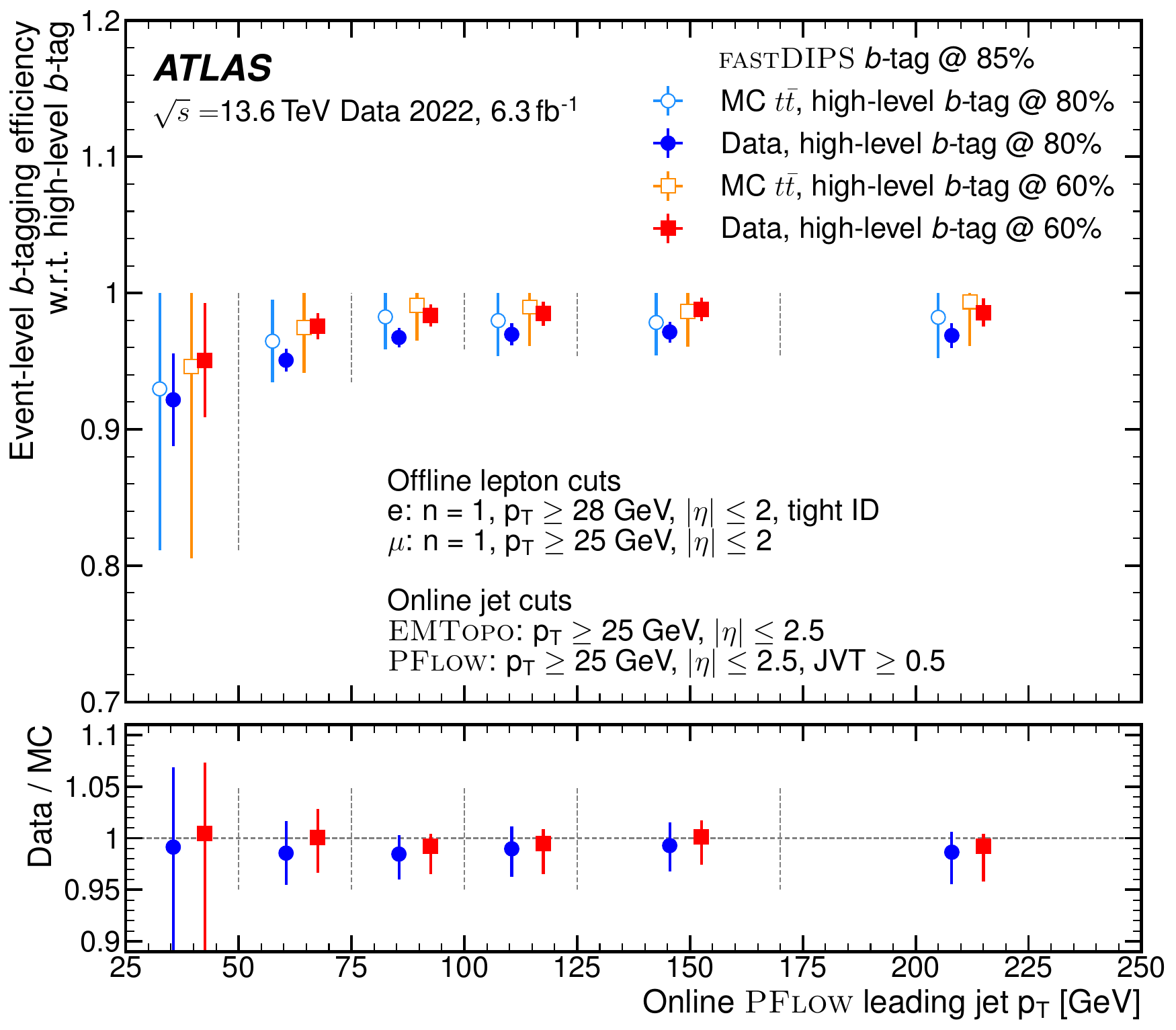}
}
\subcaptionbox{\label{fig:Jlvl_ptEff}}[0.49\textwidth]{
\includegraphics[width=\hsize]{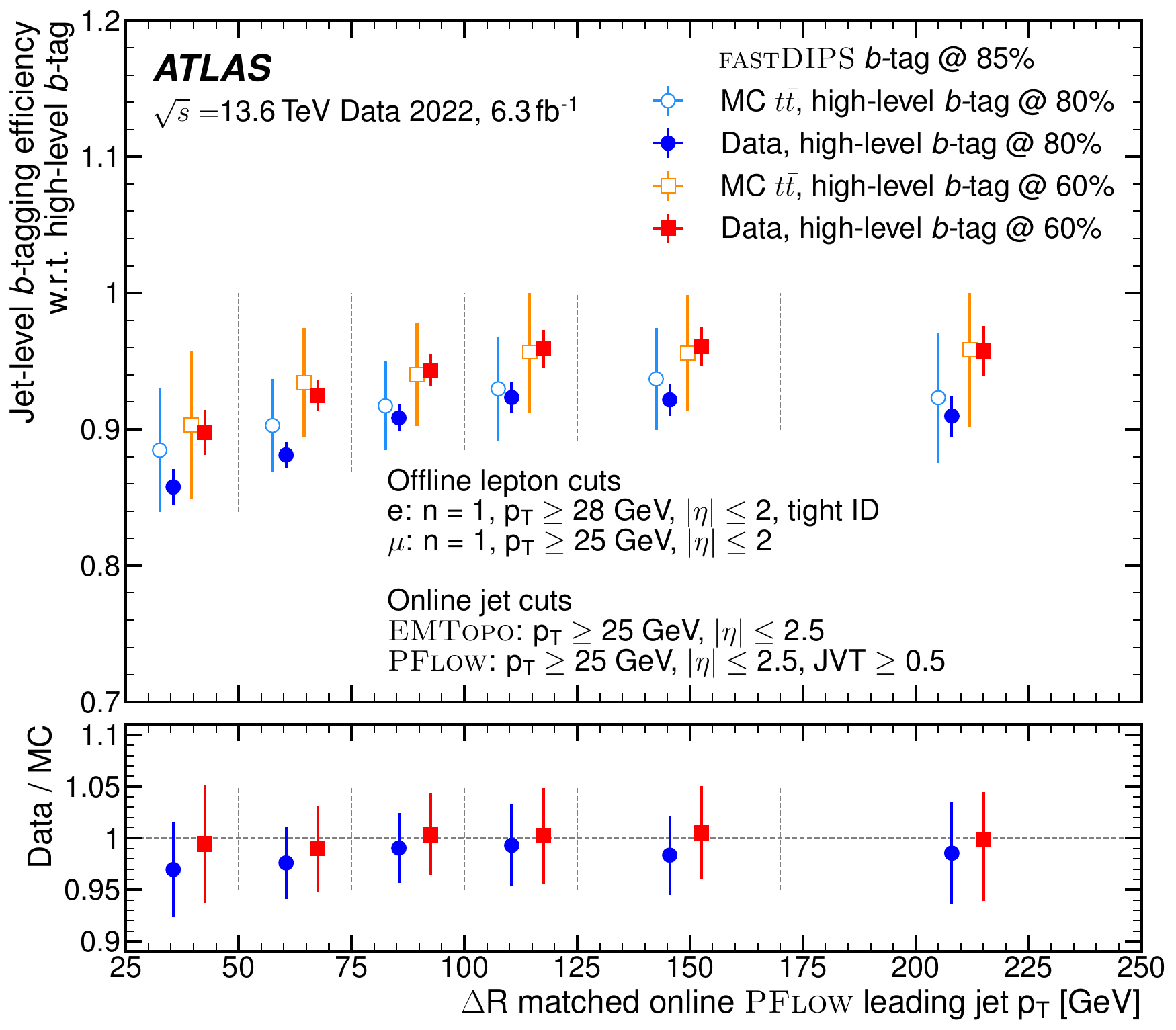}
}
\caption{\label{fig:effs}%
Event-level and jet-level efficiencies in simulated $t\bar{t}$ events and data
recorded by a trigger selecting $t\bar{t}$-like events, at $\sqrt{s}=13.6\,$\TeV.
Figure \subref{fig:Elvl_ptEff} shows the event-level efficiency for at least one fast $b$-tag, in bins of the leading jet \pt{}, in events that already have at least one \emtopo jet and one high-level $b$-tag.
Figure \subref{fig:Jlvl_ptEff} shows the corresponding jet-level efficiency for the fast $b$-tagging algorithm with respect to the high-level $b$-tagging one.
The two plots have the same binning, bin edges are displayed with the vertical dashed lines.
For the jet-level efficiency,  \emtopo and \empflow jets are geometrically matched with a $\Delta R < 0.3$, and \pt relative difference less than 10\%.
In both figures, the displayed uncertainties are statistical only. In the ratio panels, the errors are propagated as the quadratic sum of the statistical uncertainties.
}
\end{figure}

The efficiency of the fast $b$-tagging preselection with respect to the high-level $b$-tagging algorithm is
tested both at the jet- and event-level with the metrics defined in Eq.~(\ref{eq:jetEff}) and~(\ref{eq:evtEff}).
The evaluated fast $b$-tagging working point is the nominal 85\% used in the trigger selections during 2022 data taking.
The high-level $b$-tagging algorithm is used at two working points for comparison, 80\% and 60\%, both used in different trigger selections during 2022 data taking.
The event-level efficiency for each case is reported in Figure \ref{fig:effs}\subref{fig:Elvl_ptEff}: the \fastdips{} selection has an efficiency ranging from 93\% to 99\%.
The jet-level efficiencies are studied for the same working points, 85\% for the fast $b$-tagging algorithm and both 80\% and 60\% for the high-level $b$-tagging algorithm. %
The results are shown in Figure \ref{fig:effs}\subref{fig:Jlvl_ptEff}.

\begin{figure}[h]
\centering
\includegraphics[width=0.5\textwidth]{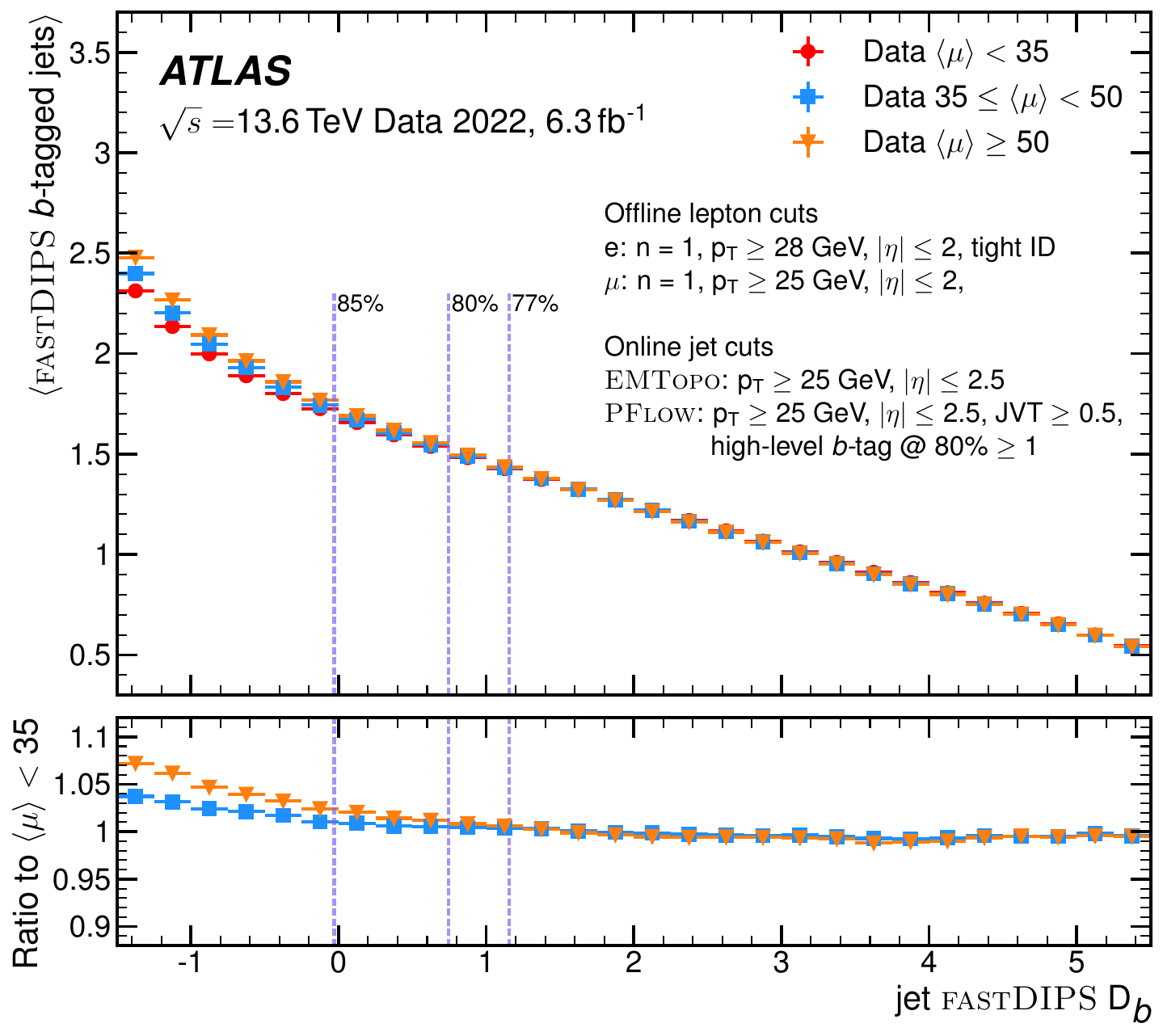}
\caption{\label{fig:avgBtagMultiplictyPerMu}
Mean number of \btagged jets per event as a function of the \fastdips{} discriminant selection employed, in $t\bar{t}$ enriched data. The means are computed in data events, binned with respect to the average number of interactions per bunch crossing ($\langle\mu\rangle$).
During the period of data taking considered, the value of $\langle\mu\rangle$ ranges from roughly 20 up to 60.
Statistical uncertainties, reported as vertical lines, are smaller than the marker sizes, and therefore not visible.
Vertical dashed lines represent three commonly used working points.}
\end{figure}

We studied the stability of \fastdips{} under different pile-up conditions by binning data with respect to the actual number of $pp$ interactions per bunch crossing, denoted by $\langle\mu\rangle$.
The trend of the average number of \btagged jets per event, for progressively more stringent \fastdips{} selections, is shown in three different bins of $\langle\mu\rangle$ in Figure~\ref{fig:avgBtagMultiplictyPerMu}.
For high pile-up events, the \btagged jet multiplicity increases with looser network selections, due to the increased number of mis-tagged jets.
For tighter discriminant selections, such as those used for defining the \bjet working points, the pile-up dependence is reduced.

\subsection{Impact on \dihfourb{} signal acceptance}

The detection of multi-\bhadron signatures, specifically \dihfourb{}, challenges the capacity of the ATLAS trigger and data acquisition system by requiring multiple \btagged jets at low \pt, in a region of phase space that is overwhelmed by light jets from QCD interactions.
HL-LHC trigger studies~\cite{ATLAS-TDR-29} have shown that raising \pT{} requirements will reduce trigger rates to an acceptable level, but at a significant cost in \dihfourb{} acceptance.
In Run~3, to maximise Standard Model \dihfourb{} signal acceptance, ATLAS relies on a relatively high rate ($8\,\text{kHz}$) L1 seed.
The fast \btagging algorithm reduces this rate further, to a level that is affordable in the current HLT CPU farm.

It is essential to verify that the gain achieved by the high-rate L1 item is maintained despite the introduction of the fast \btagging preselection.
The impact of that preselection is demonstrated in Table~\ref{tab:fast_performance} in terms of both rejection power and physics acceptance, using \dihfourb{} as a benchmark channel.
The rejection factors of the HLT fast \btagging preselection on top of L1 are estimated with Run~2 enhanced bias data~\cite{ATL-DAQ-PUB-2016-002}, which are representative of the data that passes the L1 trigger.
The L1 selection requires three central ($|\eta| < 2.5$) jets with $\ET>15$~\GeV, out of which the leading one is required to have $\ET>45$~\GeV{} and $|\eta| < 2.1$.
The impact on the \dihfourb{}  acceptance is evaluated in an HLT selection that requires four \empflow jets ($\pT > \{80, 55, 28, 20\}\,\GeV$), two of which must be \btagged with a high-level $b$-tagging discriminant above the 77\% efficiency threshold~\cite{FTAG-2018-01}.
Preselection with fast \btagging requires at least four \emtopo jets with $\pt > 20\,\GeV$ (which may overlap with the \empflow jets in the high-level \btagging), two of which must pass a \fastdips{} requirement.

The HLT fast \btagging preselection reduces the rate at which the high-level algorithm is running by a factor up to ten, which is necessary to reduce the L1 rate enough for the event-wide tracking to run at the HLT at the peak luminosity of the Run~3 data taking ($2\times 10^{34}$ cm$^{-2}$s$^{-1}$).
It also reduces the HLT rate by a small factor (around 10\%).
The relative trigger acceptance is calculated for \dihfourb{} simulated events, comparing the HLT fast \btagging preselection to the HLT \dihfourb{} selection described above. We observed a negligible loss of \dihfourb{} acceptance from the fast \bjet preselection, at a level of 2--4\%.
This acceptance loss is dwarfed by the much larger loss that would result from raised jet $\pt$ thresholds or random trigger vetoes, at least one of which would be required without the \fastdips{} preselection.

\newcolumntype{M}[1]{>{\centering}m{#1}}

\begin{table}
\centering
\begin{tabular}{ | >{\centering\arraybackslash}m{6cm} || >{\centering\arraybackslash}m{3.5cm}  | >{\centering\arraybackslash}m{3.5cm} |} \hline
\centering{Trigger selection} & \centering{Preselection rejection \newline factor on top of L1} &  \dihfourb{} relative trigger acceptance \\ \hline \hline
\centering{L1 + HLT preselection (85\% WP) + HLT selection ($HH\rightarrow b\bar{b}b\bar{b}$)} & \centering{$\sim$ 5} &   0.98 \\ \hline
\centering{L1 + HLT preselection (80\% WP) + HLT selection ($HH\rightarrow b\bar{b}b\bar{b}$)} & \centering{$\sim$ 10} &  0.96 \\ \hline
\end{tabular} \\ \vspace{0.1cm}
\caption{%
\label{tab:fast_performance}
Fast \btagging trigger selection performance and impact on the \dihfourb{} acceptance. Here \newterm{preselection} requires four \emtopo jets with $\pt > 20\,\GeV$, two of which must be \btagged with \fastdips. The acceptance  for the \dihfourb{} signature of the trigger selection that is summarised in the first column of the table is evaluated relative to a trigger selection that does not include the HLT preselection.
}
\end{table}


%
%
%

%
\section{Conclusion}
\label{sec:conclusion}
%

The ATLAS high-level trigger has improved significantly in Run~3, in part owing to the use of tracking and other offline-like algorithms in hadronic signatures.
These improvements come with a high CPU cost, and taken alone would have increased the CPU demands beyond what would be feasible in the ATLAS trigger computing farm.
Flavour tagging, used in multi-$b$ selections and essential for signatures such as \dihfourb{}, was among the dominant contributors to these demands.
This aspect was mitigated through the introduction of a fast \btagging preselection, which runs after calorimeter-based jet finding and before the much more expensive full-event track reconstruction.
By restricting tracking to limited regions of interest surrounding energetic jets, this preselection serves as a crude and fast veto.
This veto reduces CPU load from multi-$b$ signatures and brings the total resource use safely within the computing farm's budget.
Thanks to this development, ATLAS is able to save  \dihfourb{} events at a higher rate than ever before.

The method outlined in this paper has tremendous potential for application at the HL-LHC, where projections have shown that raised jet $\pt$ thresholds substantially lower the experiment's sensitivity to \dihfourb.
With more challenging pile-up conditions, tracking will become even more difficult to execute.
New approaches such as the one presented here, which ensure high physics acceptance rates while optimising resource use, are imperative for attaining the ambitious physics goals of the ATLAS experiment.


%
\section*{Acknowledgements}
%

%
%

%
%

We thank CERN for the very successful operation of the LHC, as well as the
support staff from our institutions without whom ATLAS could not be
operated efficiently.

We acknowledge the support of
ANPCyT, Argentina;
YerPhI, Armenia;
ARC, Australia;
BMWFW and FWF, Austria;
ANAS, Azerbaijan;
CNPq and FAPESP, Brazil;
NSERC, NRC and CFI, Canada;
CERN;
ANID, Chile;
CAS, MOST and NSFC, China;
Minciencias, Colombia;
MEYS CR, Czech Republic;
DNRF and DNSRC, Denmark;
IN2P3-CNRS and CEA-DRF/IRFU, France;
SRNSFG, Georgia;
BMBF, HGF and MPG, Germany;
GSRI, Greece;
RGC and Hong Kong SAR, China;
ISF and Benoziyo Center, Israel;
INFN, Italy;
MEXT and JSPS, Japan;
CNRST, Morocco;
NWO, Netherlands;
RCN, Norway;
MEiN, Poland;
FCT, Portugal;
MNE/IFA, Romania;
MESTD, Serbia;
MSSR, Slovakia;
ARRS and MIZ\v{S}, Slovenia;
DSI/NRF, South Africa;
MICINN, Spain;
SRC and Wallenberg Foundation, Sweden;
SERI, SNSF and Cantons of Bern and Geneva, Switzerland;
MOST, Taiwan;
TENMAK, T\"urkiye;
STFC, United Kingdom;
DOE and NSF, United States of America.
In addition, individual groups and members have received support from
BCKDF, CANARIE, Compute Canada and CRC, Canada;
PRIMUS 21/SCI/017 and UNCE SCI/013, Czech Republic;
COST, ERC, ERDF, Horizon 2020 and Marie Sk{\l}odowska-Curie Actions, European Union;
Investissements d'Avenir Labex, Investissements d'Avenir Idex and ANR, France;
DFG and AvH Foundation, Germany;
Herakleitos, Thales and Aristeia programmes co-financed by EU-ESF and the Greek NSRF, Greece;
BSF-NSF and MINERVA, Israel;
Norwegian Financial Mechanism 2014-2021, Norway;
NCN and NAWA, Poland;
La Caixa Banking Foundation, CERCA Programme Generalitat de Catalunya and PROMETEO and GenT Programmes Generalitat Valenciana, Spain;
G\"{o}ran Gustafssons Stiftelse, Sweden;
The Royal Society and Leverhulme Trust, United Kingdom.

The crucial computing support from all WLCG partners is acknowledged gratefully, in particular from CERN, the ATLAS Tier-1 facilities at TRIUMF (Canada), NDGF (Denmark, Norway, Sweden), CC-IN2P3 (France), KIT/GridKA (Germany), INFN-CNAF (Italy), NL-T1 (Netherlands), PIC (Spain), ASGC (Taiwan), RAL (UK) and BNL (USA), the Tier-2 facilities worldwide and large non-WLCG resource providers. Major contributors of computing resources are listed in Ref.~\cite{ATL-SOFT-PUB-2021-003}.

%
%


%
%
%
%
%
%
%

%
%
%
%
%
%

%
%

%
%
%
\printbibliography

\clearpage
 
\begin{flushleft}
\hypersetup{urlcolor=black}
{\Large The ATLAS Collaboration}

\bigskip

\AtlasOrcid[0000-0002-6665-4934]{G.~Aad}$^\textrm{\scriptsize 102}$,
\AtlasOrcid[0000-0002-5888-2734]{B.~Abbott}$^\textrm{\scriptsize 120}$,
\AtlasOrcid[0000-0002-1002-1652]{K.~Abeling}$^\textrm{\scriptsize 55}$,
\AtlasOrcid[0000-0001-5763-2760]{N.J.~Abicht}$^\textrm{\scriptsize 49}$,
\AtlasOrcid[0000-0002-8496-9294]{S.H.~Abidi}$^\textrm{\scriptsize 29}$,
\AtlasOrcid[0000-0002-9987-2292]{A.~Aboulhorma}$^\textrm{\scriptsize 35e}$,
\AtlasOrcid[0000-0001-5329-6640]{H.~Abramowicz}$^\textrm{\scriptsize 151}$,
\AtlasOrcid[0000-0002-1599-2896]{H.~Abreu}$^\textrm{\scriptsize 150}$,
\AtlasOrcid[0000-0003-0403-3697]{Y.~Abulaiti}$^\textrm{\scriptsize 117}$,
\AtlasOrcid[0000-0003-0762-7204]{A.C.~Abusleme~Hoffman}$^\textrm{\scriptsize 137a}$
% ...
\bigskip
\\

% ...
$^{29}$Physics Department, Brookhaven National Laboratory, Upton NY; United States of America.\\
% ...
$^{35}$$^{(a)}$Facult\'e des Sciences Ain Chock, R\'eseau Universitaire de Physique des Hautes Energies - Universit\'e Hassan II, Casablanca;$^{(b)}$Facult\'{e} des Sciences, Universit\'{e} Ibn-Tofail, K\'{e}nitra;$^{(c)}$Facult\'e des Sciences Semlalia, Universit\'e Cadi Ayyad, LPHEA-Marrakech;$^{(d)}$LPMR, Facult\'e des Sciences, Universit\'e Mohamed Premier, Oujda;$^{(e)}$Facult\'e des sciences, Universit\'e Mohammed V, Rabat;$^{(f)}$Institute of Applied Physics, Mohammed VI Polytechnic University, Ben Guerir; Morocco.\\
% ...
$^{49}$Fakult\"{a}t Physik , Technische Universit{\"a}t Dortmund, Dortmund; Germany.\\
% ...
$^{55}$II. Physikalisches Institut, Georg-August-Universit\"{a}t G\"ottingen, G\"ottingen; Germany.\\
% ...
$^{102}$CPPM, Aix-Marseille Universit\'e, CNRS/IN2P3, Marseille; France.\\
% ...
$^{117}$Department of Physics, New York University, New York NY; United States of America.\\
% ...
$^{120}$Homer L. Dodge Department of Physics and Astronomy, University of Oklahoma, Norman OK; United States of America.\\
% ...
$^{137}$$^{(a)}$Departamento de F\'isica, Pontificia Universidad Cat\'olica de Chile, Santiago;$^{(b)}$Millennium Institute for Subatomic physics at high energy frontier (SAPHIR), Santiago;$^{(c)}$Instituto de Investigaci\'on Multidisciplinario en Ciencia y Tecnolog\'ia, y Departamento de F\'isica, Universidad de La Serena;$^{(d)}$Universidad Andres Bello, Department of Physics, Santiago;$^{(e)}$Instituto de Alta Investigaci\'on, Universidad de Tarapac\'a, Arica;$^{(f)}$Departamento de F\'isica, Universidad T\'ecnica Federico Santa Mar\'ia, Valpara\'iso; Chile.\\
% ...
$^{150}$Department of Physics, Technion, Israel Institute of Technology, Haifa; Israel.\\
$^{151}$Raymond and Beverly Sackler School of Physics and Astronomy, Tel Aviv University, Tel Aviv; Israel.\\
% ...
$^{a}$ Also Affiliated with an institute covered by a cooperation agreement with CERN.\\
$^{b}$ Also at An-Najah National University, Nablus; Palestine.\\
$^{c}$ Also at APC, Universit\'e Paris Cit\'e, CNRS/IN2P3, Paris; France.\\
$^{d}$ Also at Borough of Manhattan Community College, City University of New York, New York NY; United States of America.\\
$^{e}$ Also at Center for High Energy Physics, Peking University; China.\\
$^{f}$ Also at Center for Interdisciplinary Research and Innovation (CIRI-AUTH), Thessaloniki ; Greece.\\
% ...

\end{flushleft}


\end{document}